\documentclass[aps,prc,superscriptaddress,showpacs,floatfix,twocolumn,preprintnumbers]{revtex4}
\usepackage{graphicx}
\usepackage{dcolumn}
\usepackage{amsmath,amssymb,bm}

\newcommand{\beq}{\begin{equation}}
\newcommand{\eeq}{\end{equation}}
\newcommand{\bea}{\begin{eqnarray}}
\newcommand{\eea}{\end{eqnarray}}
\newcommand{\Order}{{\cal O}}
\newcommand{\no}{\nonumber}
\newcommand{\Lagr}{{\cal L}}
\newcommand{\sh}{\!\not\!}
\newcommand{\Sh}{\!\not\!\!}

\newcommand{\Fv} [1]{{F_{#1}^{\!\not\,v}}}
\newcommand{\Fs} [1]{{F_{#1}^{\!\not\,s}}}
\newcommand{\Fvs}[1]{{F_{#1}^{\!\not\,v,\not\,s}}}

\newcommand{\bFv} [1]{{\bar F_{#1}^{\!\not\,v}}}
\newcommand{\bFs} [1]{{\bar F_{#1}^{\!\not\,s}}}
\newcommand{\bFvs}[1]{{\bar F_{#1}^{\!\not\,v,\not\,s}}}

\newcommand{\Gv} [1]{{G_{#1}^{\!\not\,v}}}
\newcommand{\Gs} [1]{{G_{#1}^{\!\not\,s}}}
\newcommand{\Gvs}[1]{{G_{#1}^{\!\not\,v,\not\,s}}}

\newcommand{\kv} {{\kappa^{\!\not\,v}}}
\newcommand{\ks} {{\kappa^{\!\not\,s}}}
\newcommand{\kvs}{{\kappa^{\!\not\,v,\not\,s}}}

\newcommand{\rv} [1]{{\rho_{#1}^{\!\not\,v}}}
\newcommand{\Rv} [1]{\bigl(\rho_{#1}^{\!\not\,v}\bigr)}
\newcommand{\rs} [1]{{\rho_{#1}^{\!\not\,s}}}
\newcommand{\rvs}[1]{{\rho_{#1}^{\!\not\,v,\not\,s}}}
\newcommand{\rud}[1]{{\rho_{#1}^{u,d}}}
\newcommand{\Rud}[1]{\bigl(\rho_{#1}^{u,d}\bigr)}

\begin{document}

\preprint{HISKP-TH-06/10}

\title{Isospin violation in the vector form factors of the nucleon}

\author{Bastian Kubis}
\email{kubis@itkp.uni-bonn.de}
\affiliation{Helmholtz-Institut f\"ur Strahlen- und Kernphysik (Theorie),
             Universit\"at Bonn, Nussallee 14-16, 53115 Bonn, Germany.}

\author{Randy Lewis}
\email{randy.lewis@uregina.ca}
\affiliation{Department of Physics, University of Regina, Regina,
             Saskatchewan, Canada, S4S 0A2.}

\date{\today}

\begin{abstract}
A quantitative understanding of isospin violation
is an increasingly important ingredient for the 
extraction of the nucleon's strange vector form factors
from experimental data.  We calculate the isospin
violating electric and magnetic form factors in
chiral perturbation theory to leading and
next-to-leading order respectively, and we extract
the low-energy constants from resonance saturation.
Uncertainties are dominated largely by limitations
in the current knowledge of some vector meson couplings.
The resulting bounds on isospin violation are
sufficiently precise to be of value to on-going
experimental studies of the strange form factors.
\end{abstract}


\maketitle

\section{Introduction}

Since a nucleon has no valence strange quarks, the strange form factors
provide information about the dynamics of virtual quarks within a nucleon.
In recent years the strange electric and magnetic form factors have been
studied experimentally by the SAMPLE~\cite{SAMPLE}, A4~\cite{A4a,A4b}, 
HAPPEX~\cite{HAPPEXa,HAPPEXb}, and G0~\cite{G0}
Collaborations, and theoretically via 
lattice QCD simulations~\cite{Dong,Mathur,LewisWilcox,Lein1,Lein2,Lein3},
chiral perturbation theory~\cite{MusolfIto,HMS,HKM,HammerPuglia,KubisPAVI},
and hadron models~\cite{Silva,Silva:2005qm,Jido,Carvalho,Chen,Bijker,An,Riska}.
(For a discussion of earlier theoretical studies, see Ref.~\cite{Ramsey-Musolf}.)
A central conclusion of recent research is that the strange electric and
magnetic form factors are small, perhaps even small enough to require a
careful understanding of the competing effects from isospin violation.

The inequality of up and down quarks, in terms of both mass and electric
charge, produces effects that mimic the strange form factors.
Extraction of the authentic strange quark effects from experimental data
requires this isospin violation to be understood theoretically.
There have been a number of theoretical studies of isospin
violation in this 
context~\cite{Pollock,Miller,LewisMobed,Ma,Thomas},
but it is not yet clear that uncertainties in these results are negligible with
respect to the strange form factors.  
Refs.~\cite{Pollock,Miller} used quark model discussions
to determine the isospin violating contributions, and found them to be less
than 1\%; in fact the strange magnetic form factor received no contribution
at all for vanishing momentum transfer.  This null contribution at $Q^2=0$
is not due to any symmetry of QCD,
and therefore the
chiral perturbation theory (ChPT) study in Ref.~\cite{LewisMobed} 
did find a non-zero
contribution, but it includes an undetermined low-energy constant.
One could think of extracting
the physics of this constant, for example, from the
light-cone baryon model of Ref.~\cite{Ma} 
or from the quark model approach of Ref.~\cite{Gardner}, 
though it is difficult
to see how to match these models to the consistent low-energy effective 
theory description of ChPT in order to avoid double counting problems.

In the present work we address two main goals.
First we revisit the calculation of chiral loops, this time using both a
manifestly covariant formalism~\cite{Becher} and heavy-baryon ChPT.
We find that the two formalisms offer complementary advantages for various
aspects of this calculation.  Since the isospin violating effects begin at
rather high orders in the chiral expansion, leading order (LO) $= \Order(p^4)$ and
next-to-leading order (NLO) $=\Order(p^5)$ (where $p$ collectively stands for
small parameters like momenta, the pion mass, or the electric coupling),
the cross-check of the entire calculation in both formalisms is valuable, and
we are able to identify and correct some errors in Ref.~\cite{LewisMobed}.

Our second major goal in the present work is to estimate the lone
combination of low-energy
constants that appears in the chiral perturbation theory expressions.
For this, we employ resonance saturation.  The physics producing isospin
violation in the nucleon's vector form factors is seen to be $\rho$--$\omega$ mixing,
and numerical values for the required vector meson couplings are obtained from
recent dispersive analyses.

A question that cannot be fully answered by our work relates
to the convergence of the chiral expansion.  With only LO and NLO results,
it is impossible to make any definitive statement about convergence.
A completely consistent extension beyond NLO seems unwarranted, since it would involve two-loop
integrals and would introduce additional unknown low-energy constants.
However, given the apparent smallness of the strange form factors and of the
isospin violating form factors, we can neglect contributions that
are simultaneously isospin violating and strange.  This allows
the discussion of isospin violation to occur within two-flavor
ChPT, where convergence properties are known to be dramatically
better than for the three-flavor theory.

The present work thus provides a complete discussion of the physics
at LO for the electric and at NLO for the magnetic isospin violating form
factors, producing numerical
values for isospin violation in the magnetic moment, electric radius,
and magnetic radius, which are of direct relevance to the ongoing experimental
studies of strange form factors.

The outline of our presentation is as follows.
We define the form factors under consideration and their leading moments
in Section~\ref{sec:basics}.
The chiral representation of the Dirac form factors to leading order
and of the Pauli form factors to next-to-leading order is (re-)derived
in some detail in Section~\ref{sec:ChPT}.  
In order to provide resonance-saturation estimates for unknown
chiral low-energy constants, we calculate the contributions
of $\rho$--$\omega$ mixing in Section~\ref{sec:res}, which 
also allow for some insight into higher-order contributions beyond
what is strictly calculated in ChPT.
The pertinent formalism to do this consistently with chiral constraints
is included.
We conclude our findings in Section~\ref{sec:Conclusions}.
Some technical details are relegated to the appendices.

\section{Basic Definitions\label{sec:basics}}

We define the isospin symmetry breaking Dirac and Pauli form factors
according to
\beq \begin{split}
\bigl\langle p(\vec p\,')&+n(\vec p\,')\,\bigl|\,
   \frac{1}{2}\bigl(\bar u\gamma_\mu u-\bar d\gamma_\mu d\bigr)
                        \,\bigr|\,p(\vec p)+n(\vec p)\bigr\rangle
\\ & 
=~ \bar u(p') \Bigl[ \gamma_\mu \Fv{1}(t)
+ \frac{i\sigma_{\mu\nu}q^\nu}{2m_N} \Fv{2}(t) \Bigr] u(p) ~, 
\\  
\bigl\langle p(\vec p\,')&-n(\vec p\,')\,\bigl|\,
   \frac{1}{6}\bigl(\bar u\gamma_\mu u+\bar d\gamma_\mu d\bigr)
                        \,\bigr|\,p(\vec p)+n(\vec p)\bigr\rangle
\\ & 
=~ \bar u(p') \Bigl[ \gamma_\mu \Fs{1}(t)
+ \frac{i\sigma_{\mu\nu}q^\nu}{2m_N} \Fs{2}(t) \Bigr] u(p) ~,
\end{split} \label{defFvs} \eeq 
where $q_\mu=p'_\mu-p_\mu$, $t=q^2$.
$\Sh\! v$, $\Sh\! s$ refer to isospin breaking pieces
in the isovector and isoscalar vector currents, respectively.
Our notation is linked to that used in Refs.~\cite{Pollock,LewisMobed} by
\[
\Fv{i}(t) = {{}^{u-d}F_i^{p+n}}(t) ~, \quad
\Fs{i}(t) = {{}^{u+d}F_i^{p-n}}(t) ~.
\]
The Sachs form factors are defined in terms of these as
\beq \begin{split}
\Gvs{E}(t) &= 
\Fvs{1}(t) + \frac{t}{4m_N^2} \Fvs{2}(t) ~,\\
\Gvs{M}(t) &= 
\Fvs{1}(t) + \Fvs{2}(t) ~.
\end{split} \label{Sachs}\eeq
Current conservation implies $\Fvs{1}(0)=0$,
while the Pauli form factors are normalized to the 
isospin breaking pieces of the 
(anomalous) magnetic moments,
\beq
\Fvs{2}(0) ~=~ \kvs ~.
\eeq
To complete the definition of the leading moments of these form factors,
we define radius-like terms as the coefficients of form factor terms linear in $t$,
\beq \begin{split}
\Fvs{1/2}(t) &=
\Fvs{1/2}(0) + \rvs{1/2}\,t
+\Order(t^2) ~, \\ 
\Gvs{E/M}(t) &=
\Gvs{E/M}(0) + \rvs{E/M}\,t
+\Order(t^2) ~. 
\end{split} \eeq
Finally, the proton's neutral weak form factors $G_{E/M}^{p,Z}$
depend on one  specific linear
combination of isospin breaking form factors,
\beq \begin{split}
G_{E/M}^{p,Z}(t) =&\, \bigl(1-4\sin^2\theta_W \bigr) G_{E/M}^p(t) - G_{E/M}^n(t) \\
& - G_{E/M}^s(t) - G_{E/M}^{u,d}(t) ~, \label{GpZ}
\end{split}\eeq
where $\theta_W$ is the weak mixing angle, 
$G_{E/M}^s(t)$ are the strange vector form factors, and
\beq
G_{E/M}^{u,d}(t) \doteq \Gs{E/M}(t) - \Gv{E/M}(t) ~.
\eeq
We will use an analogous notation also for the leading moments, i.e.\ 
$\kappa^{u,d} = \ks - \kv$ etc.
Eq.~\eqref{GpZ} demonstrates that the isospin breaking form factors
``simulate'' strangeness form factors even in a world without 
strange quarks~\cite{LewisMobed,LewisPiN}, such that they have
a direct impact on the accuracy to which the strangeness form factors
can be extracted from data.

\section{Chiral Perturbation Theory\label{sec:ChPT}}

\subsection{Effective Lagrangians\label{sec:Lagr}}

We have performed the relevant loop calculations both 
in the heavy-baryon formalism~\cite{Jenkins,BKMreview}
and in the infrared regularization scheme of relativistic baryon 
ChPT~\cite{Becher}.  For reasons of briefness, we only
display the relativistic forms of the relevant Lagrangian terms 
in this section.  The equivalent heavy-baryon forms can be found in 
the quoted references.

The Goldstone boson (pion) Lagrangian is given, at leading order, by
\beq
\Lagr_{\pi\pi}^{(2)} = \frac{F^2}{4}\langle u_\mu u^\mu + \chi_+\rangle ~,
\eeq
where the matrices $u^2=U$ collect the pion fields in the usual manner,
$u_\mu = i u^\dagger \nabla_\mu U u^\dagger$ with the covariant derivative
$\nabla_\mu$.  The field $\chi_+ = u^\dagger\chi u^\dagger + u \chi^\dagger u$
contains the quark mass terms, $\chi = 2B \,\text{diag}(m_u,m_d)+\ldots$,
where $B$ is linked to the quark condensate in the chiral limit,
and $F$ can be identified with the pion decay constant, 
$F=F_\pi=92.4$~MeV. $\langle \ldots \rangle$ denotes the trace in flavor space.

A striking observation is the fact that the leading term generating the 
charged-to-neutral pion mass difference~\cite{EGPdR},
\[
\Lagr_{\pi\pi\gamma^*}^{(2)}
= Z F^4 \langle Q_+^2 - Q_-^2 \rangle 
~\Rightarrow~ M_{\pi^+}^2-M_{\pi^0}^2 = 2Ze^2F^2 ~,
\]
where $Q_\pm = \frac{1}{2}(u Q u^\dagger \pm u^\dagger Q u)$, 
$Q=\text{diag}(e,0)$,
does \emph{not} feature in the following calculation:
in nuclear physics language,
this operator only breaks charge \emph{independence}
(independence under completely general rotations in isospin space), 
but not charge \emph{symmetry} (rotations by $\pi$ about a fixed axis in
isospin space, resulting in the simultaneous exchange
of $u$ and $d$ quarks, as well as protons and neutrons)
as required for the form factors in Eq.~\eqref{defFvs}, 
so the pion mass difference alone will never generate
contributions to the form factors under investigation, and we will
neglect terms that are of second order in isospin breaking.

The leading-order pion-nucleon Lagrangian is given by
\beq
\Lagr_{\pi N}^{(1)} = \bar\Psi \Bigl\{ i \Sh D -m+\frac{g_A}{2} \sh u\gamma_5 \Bigr\} \Psi ~,
\label{L1}
\eeq
where $D_\mu$ is the usual covariant derivative acting on the nucleon,
$m$ is the nucleon mass in the chiral limit, 
and $g_A$ can be identified with the axial coupling constant, $g_A=1.26$. 
Of the second-order pion-nucleon Lagrangian (that also contains
all possible effects of virtual photons), we only quote the terms
relevant for our calculation:
\beq \begin{split}
\Lagr_{\pi N\gamma^*}^{(2)} = \bar\Psi \Bigl\{ &
c_5 \tilde\chi_+ + f_2 F^2 \langle Q_+ \rangle Q_+  \\
& +\frac{\sigma^{\mu\nu}}{8m} \bigl(
c_6 F^+_{\mu\nu} + c_7 \langle F^+_{\mu\nu} \rangle \bigr)
\Bigr\} \Psi ~.
\label{L2}
\end{split} \eeq
Here we have introduced the notation $\tilde O = O - \frac{1}{2}\langle O \rangle$
for traceless operators.
$F_{\mu\nu}^+=u^\dagger F_{\mu\nu} u + u F_{\mu\nu} u^\dagger$
contains the usual electromagnetic field strength tensor.
There are two types of terms in Eq.~\eqref{L2}:
$c_6$ and $c_7$ are related to the isovector and isoscalar anomalous
magnetic moments (in the chiral limit) according to
\beq
\kappa^v = \kappa_p-\kappa_n = c_6 ~, \quad 
\kappa^s = \kappa_p+\kappa_n = c_6+2c_7 ~,
\eeq
with the experimental values $\kappa^v =3.706$, $\kappa^s =-0.120$, 
while the operators proportional to $c_5$ and $f_2$ generate the leading 
proton-neutron mass difference
\beq
\Delta m \doteq m_n-m_p = 4B(m_u-m_d)c_5+e^2F^2f_2 ~.
\eeq
As all our results will be expressible in terms of $\Delta m$,
we do not have to care about the precise separation of this mass difference
into its strong ($\propto c_5$) and electromagnetic ($\propto f_2$) parts.
We have neglected to spell out terms in Eq.~\eqref{L2} 
that lead only to a common nucleon mass shift and can be taken care of
by replacing the nucleon mass in the chiral limit by its physical value,
$m \to m_N = \frac{1}{2}(m_p+m_n) $.

Finally, the following fourth-order terms~\cite{LpiN4,LpiNgamma} 
contribute to the isospin breaking form factors:
\beq \begin{split}
\Lagr_{\pi N{\gamma^*}}^{(4)} &= 
\bar\Psi \Bigl\{ -\frac{e_{107}}{2} \langle F_{\mu\nu}^+ \rangle \tilde\chi_+
-\frac{e_{108}}{2} \langle \tilde F_{\mu\nu}^+ \tilde\chi_+ \rangle \\
&\qquad +
h_{40} F^2\langle Q_+ \rangle \langle \tilde Q_+ \tilde F_{\mu\nu}^+ \rangle \\
&\qquad + 
h_{43} F^2\langle Q_+ \rangle  \tilde Q_+ \langle F_{\mu\nu}^+\rangle
\Bigr\} \sigma^{\mu\nu}\Psi \,.
\end{split}\label{L4} 
\eeq
The terms proportional to $e_{107}$, $h_{43}$ contribute to 
$\ks$, the terms multiplied by $e_{108}$, $h_{40}$ to
$\kv$.  Therefore, both form factors contain one counterterm 
each proportional to $m_u-m_d$ and $e^2$.  
The constants $e_{108}$, $h_{40}$ contain a divergent piece,
\[
e_{108} = e^r_{108}(\lambda) + \beta_{e_{108}} L ~, \quad
h_{40}  = h^r_{40}(\lambda)  + \beta_{h_{40}}  L ~,
\]
where $L$ has a pole in $(d-4)$,
\[
L = \frac{\lambda^{d-4}}{16\pi^2} \biggl\{
\frac{1}{d-4} + \frac{1}{2}\bigl[ \gamma_E -1-\log 4\pi \bigr] \biggr\} ~,
\]
$\lambda$ is the dimensional regularization scale,
and $\gamma_E$ is Euler's constant.
The scale-dependent finite constants $e^r_{108}$, $h^r_{40}$
obey a renormalization group equation according to
\[
\lambda \frac{d}{d\lambda} e^r_{108}(\lambda) = - \frac{\beta_{e_{108}}}{16\pi^2} ~, \quad
\lambda \frac{d}{d\lambda} h^r_{40} (\lambda) = - \frac{\beta_{h_{40} }}{16\pi^2} ~.
\]
In order to render the isospin breaking magnetic moments finite
and scale-independent, we find the 
beta functions
\[
\beta_{e_{108}} = \frac{2g_A^2c_5}{F^2} ~, \qquad
\beta_{h_{40} } = -\frac{g_A^2f_2}{F^2} ~.
\]
In contrast, $e_{107}$ and $h_{43}$ are finite and scale-independent.

It should be mentioned that Ref.~\cite{LewisMobed} included the
$\Delta(1232)$ as an explicit field in ChPT.  When the
corresponding loop diagrams were computed, their
momentum dependences were found to be negligible
relative to the final error bands on the form factors.
Since these effects would also be insignificant relative to
the error bands that will be obtained in the present work,
we omit the explicit $\Delta(1232)$ field from the outset.
In the framework used here, its effects would show up 
via higher-order low-energy constants that are beyond the
accuracy we aim at in this work.

\subsection{Power counting\label{sec:power}}

We repeat here some power counting arguments that were mostly already
put forward in Ref.~\cite{LewisMobed}.

It is easy to see that the usual (isospin conserving) vector form factors 
receive polynomial contributions to the leading moments,
i.e.\ magnetic moment, electric (or Dirac) radius, magnetic (or Pauli)
radius, at chiral orders $p^2$, $p^3$, $p^4$, respectively.  
Pion loop contributions start at $\Order(p^3)$, 
therefore such loop effects can generate a parameter-free
leading-order prediction for the magnetic radius.
The leading magnetic radius term has to scale like $M_\pi^{-1}$; 
this infrared singularity has been well-known for a long time~\cite{Beg}.

\begin{figure}
\includegraphics[height=4.2cm]{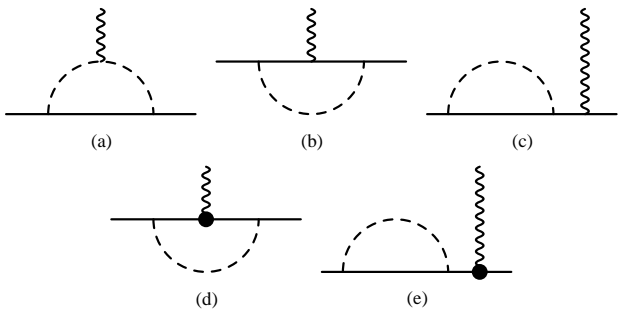}
\caption{\label{fig:pionloops} 
One-pion-loop diagrams. 
Full/dashed/wiggly lines denote nucleons/pions/vector currents, respectively.
The filled circles in diagrams (d) and (e) represent the magnetic couplings
from $\Lagr_{\pi N\gamma^*}^{(2)}$. 
The crossed diagrams of (c) and (e) are not depicted separately.
Mass insertions on the nucleon propagator lines,
yielding the physical proton and neutron masses,
are not explicitly shown.
}
\end{figure}
From the previous section, it is obvious that polynomial contributions
to the leading moments of the \emph{isospin violating} form factors
always have to include factors of either $m_u-m_d$ or $e^2$, therefore
they are suppressed by two orders in the chiral power counting 
with respect to the isospin conserving ones.  
We therefore find precisely the terms of $\Lagr_{\pi N\gamma^*}^{(4)}$
in Eq.~\eqref{L4} contributing to the magnetic moments, while 
operators of electric and magnetic radius type would arise at 
orders $p^5$ and $p^6$.  
On the other hand, the pion loop diagrams with additional mass insertions
of the $c_5$, $f_2$ operators in Eq.~\eqref{L2}, 
shown in Fig.~\ref{fig:pionloops}, start to contribute
only one order higher than the same diagrams without mass corrections, 
therefore leading loop contributions are also of order $p^4$.  
This means that, while a prediction of the isospin violating
magnetic moments is hampered by the a priori unknown 
$\Lagr_{\pi N\gamma^*}^{(4)}$ counterterms, both electric and magnetic
radii can be unambiguously predicted at leading order, 
and for the magnetic radii even the next-to-leading-order corrections
are free of unknown parameters.  
The leading infrared singularity in the magnetic radius scales like
$\Delta m/M_\pi^2$, the leading electric and subleading magnetic radius
terms like $\Delta m/M_\pi$.

\begin{figure}
\includegraphics[height=2cm]{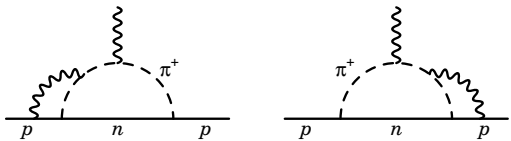}
\caption{\label{fig:photonpionloop} 
A pair of two-loop diagrams.  
The closed wiggly lines denote virtual photons.
The sum of both contributions to the magnetic form factor drops out.
}
\end{figure}
A calculation of the isospin violating Pauli (or magnetic) form factor
up to $\Order(p^5)$ is massively facilitated by the fact that no
two-loop diagrams contribute.  This can be seen as follows:
\begin{enumerate}
\item As mentioned in Section \ref{sec:Lagr}, the pion mass difference alone cannot
generate charge symmetry breaking terms.  Therefore, diagrams with
two pion loops would require another (subleading) charge symmetry breaking vertex
or the nucleon mass difference in order to contribute, which would then,
however, be at least of order $p^6$.
\item Two photon loops are of second order in isospin breaking and 
can be disregarded.  In addition, it is easily seen in the heavy-baryon
formalism that such diagrams with only leading-order photon couplings
cannot generate the spin operators necessary for a magnetic contribution.
\item Finally, there might be diagrams with one pion and one photon loop.
The only diagrams that generate a magnetic structure at $\Order(p^5)$ 
are of the type (a) in Fig.~\ref{fig:pionloops}, with one additional photon loop attached.
However, it can be checked in the heavy-baryon formulation that the 
\emph{sum} of the two diagrams in Fig.~\ref{fig:photonpionloop} is proportional to the
anticommutator of the two Pauli--Lubanski spin operators stemming
from the pion-nucleon couplings, and therefore it again only yields
a contribution to the electric form factor.  The same mechanism 
can be checked for all other possible diagrams.
\end{enumerate}
Furthermore, also one-loop diagrams with isospin-breaking vertices
(other than the nucleon mass difference insertion)
can be ruled out of consideration:
\begin{enumerate}
\item[4.] Tadpole graphs with isospin breaking couplings fail to generate
infrared singularities proportional to $M_\pi^{-1}$.
\item[5.] The third-order pion-nucleon Lagrangian contains
isospin breaking pion-nucleon coupling constants.  
However, these affect only the $\pi^0 N N$ coupling, hence they
do not contribute in a type (a) diagram,
while the remaining diagrams in Fig.~\ref{fig:pionloops}
are subleading in their contributions to the Pauli form factor
and therefore can only play a role at $\Order(p^6)$.
\end{enumerate}
\begin{figure}
\includegraphics[height=1.8cm]{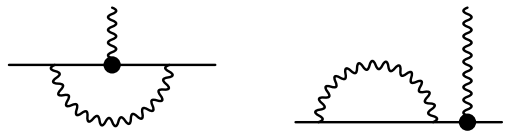}
\caption{\label{fig:photonloop}
One-photon-loop diagrams with magnetic couplings.
The sum of both diagrams vanishes.
(The addition of a crossed right diagram is implied.)}
\end{figure} 
Finally, the only one-photon loops contributing to the magnetic form factor
are the ones depicted in Fig.~\ref{fig:photonloop}.  
They have no $t$-dependence up to the chiral order considered here,
and it is easily calculated that their
contribution to $F_2(0)$ exactly cancels.  

Therefore we have proven that the only infrared singular contributions
to the charge symmetry breaking form factor radii, and all contributions
up to $\Order(p^4)$ for the Dirac and up to $\Order(p^5)$ for Pauli form
factor in addition to the $\Lagr_{\pi N\gamma^*}^{(4)}$ counterterms
in Eq.~\eqref{L4}, are given by nucleon mass difference effects in the 
diagrams in Fig.~\ref{fig:pionloops}.

\subsection{Chiral representation of the form factors\label{sec:chiralff}}

In this section, we write down the chiral representations
of the charge-symmetry breaking form factors, the Dirac form factors
to leading order, the Pauli form factors up to next-to-leading order.
We decompose all form factors according to
\beq
\Fvs{1/2}(t) \doteq \Fvs{1/2}(0)+ \bFvs{1/2}(t)
~,
\eeq
where current conservation dictates
\[
\Fv{1}(0) = \Fs{1}(0) = 0 ~.
\]
For convenience, we define an overall dimensionless pre\-factor
\beq
C = \frac{g_A^2 m_N \Delta m}{(4\pi F_\pi)^2} 
\approx 1.4\times 10^{-3}~. \label{Cdef}
\eeq
Then the $t$-independent terms of the Pauli form factors are given by
\beq \begin{split}
\Fv{2}(0) &= 4C
\biggl[ 1+2\log\frac{M_\pi}{\lambda} 
-\frac{\pi(\kappa^v+6)M_\pi}{2m_N} \biggr] \\
& +8m_N \Bigl[ e^2 F^2 h^r_{40} - 2B(m_u-m_d)e^r_{108} \Bigr]
~, \\
\Fs{2}(0) &= 
-4C\,\frac{\pi(\kappa^s+1) M_\pi}{m_N}  \\
& +8m_N \Bigl[ e^2 F^2 h_{43} - 2B(m_u-m_d)e_{107} \Bigr] ~. 
\end{split} \label{eq:kvs} \eeq
Altogether, we find for the ultimately required combination
\bea
G_M^{u,d}(0)&=&\Fs{2}(0)-\Fv{2}(0) \no\\
&=& 
-4C\bigg[1+2\log\frac{M_\pi}{\lambda} 
+\frac{\pi(2\kappa^s-\kappa^v-4)M_\pi}{2m_N} \biggr] \no\\ && 
+\kappa_{CT}^{u,d} ~, \label{eq:G0}
\eea
where
\beq
\kappa_{CT}^{u,d} = 8m_N \!\Bigl[ e^2 F^2 \bigl(h_{43}-h^r_{40}\bigr) 
                         - 2B(m_u-m_d)\bigl(e_{107}-e^r_{108}\bigr) \Bigr] .
\label{eq:kCT}
\eeq
Note that in order to ease notation, we have suppressed the scale dependence
of $h^r_{40}$, $e^r_{108}$, $\kappa_{CT}^{u,d}$ in
Eqs.~\eqref{eq:kvs}--\eqref{eq:kCT} that is necessary to compensate for 
the chiral logarithms.

Up to the order we are considering, the chiral representations 
of the form factors $\Fs{1/2}(t)$ display no $t$-dependence,
\beq
\bFs{1}(t) = \bFs{2}(t) = 0~.
\eeq
This is due to the fact that only diagram (a) in Fig.~\ref{fig:pionloops}
contains a two-pion-cut and therefore generates momentum dependence
in the low-energy region, and that the isoscalar vector current
does not couple to pions at this order.
The $t$-dependence of the form factors $\Fv{i}(t)$, 
on the other hand, is given unambiguously in terms of pion loop 
contributions and can be written as
\bea
\bFv{1}(t)
&=& -\frac{2g_A^2M_\pi\Delta m}{F_\pi^2}\Bigl[ \bar\gamma_0(t)-4\bar\gamma_3(t)\Bigr] ~,
\label{eq:F1t} \\
\bFv{2}(t)
&=& -\frac{4g_A^2m_N\Delta m}{F_\pi^2} \no \\ 
&&  \times \biggl(\xi(t) 
- \frac{M_\pi}{m_N}\Bigl[ \bar\gamma_0(t)-5\bar\gamma_3(t)\Bigr] \biggr) ~.
\label{eq:F2t}
\eea
The explicit representations of the loop functions $\xi(t)$, 
$\bar \gamma_{0/3}(t)$ are given in Appendix~\ref{app:loopfunctions}.
We note that the corresponding expressions reported in Ref.~\cite{LewisMobed}
contain errors: Eqs.~(53) and (54) in that work are too large
by factors of 2 and 4 respectively; Eqs.~\eqref{eq:F1t} and \eqref{eq:F2t} of
the present work are the correct expressions, and we have obtained
them separately from heavy-baryon ChPT and from infrared regularization.
Expanding the loop functions to linear order in $t$, we find the radius
term according to
\beq \begin{split}
\rud{M}
&= \frac{2C}{3M_\pi^2} \biggl[
1 + \frac{5\pi M_\pi}{4m_N} - \frac{ 3\pi M_\pi}{m_N} \biggr] \\
&= \frac{2C}{3M_\pi^2} \biggl[
1 - \frac{ 7\pi M_\pi}{4m_N} \biggr] ~,
\end{split} \label{eq:rMchiral}\eeq
where the first correction stems from $\Fv{1}(t)$ 
and the second from the subleading terms in $\Fv{2}(t)$.
Numerically, we find that, due to the large enhancement factor
of $7\pi/4 \approx 5.5$, the next-to-leading-order correction
reduces the leading prediction for the magnetic radius term
by as much as 80\%.

\section{Contributions from $\rho$--$\omega$ mixing\label{sec:res}}

The missing ingredient in order to make the chiral representations
of the previous section predictive is a value 
for the counterterm combination $\kappa_{CT}^{u,d}$ that is unconstrained
from symmetry arguments.
As no phenomenological information is available to fix these constants,
we have to resort to model estimates in order to get a handle on them.
The obvious model to use seems to be resonance saturation.

The method to estimate chiral low-energy constants by including
heavier resonance fields in the theory and matching their low-energy limit
to higher-order operators was first introduced systematically 
in the context of meson ChPT in Refs.~\cite{EGPdR,EGLPdR,Donoghue}.
It was shown to work rather well for the constants in $\Lagr_{\pi\pi}^{(4)}$,
in particular, those constants that receive contributions from 
vector and axial vector resonances 
were shown to be numerically dominated by these
(a modern version of the concept of ``vector meson dominance'').

In Ref.~\cite{BKMressat} this method was extended to the pion-nucleon sector,
nucleon resonances (in particular the $\Delta(1232)$) were included,
and it was demonstrated that most of the constants 
in $\Lagr_{\pi N}^{(2)}$ can also be well understood this way.
However little is known about how well this procedure works for
higher-order pion-nucleon coupling constants.  
In Ref.~\cite{Kubis} it was found for the nucleon vector form factors
that in particular the isovector couplings are not well saturated
by the $\rho$ contribution alone.  This observation agrees with
the results of various dispersive analyses of these form factors
(see e.g.\ Refs.~\cite{Hohler,MMD,HM})
that always require several resonances per channel in order to 
describe the data adequately.

Despite this rather pessimistic view, there is reason to 
believe that resonance saturation for the isospin violating 
vector form factors
might in principle work better than for the isospin 
conserving ones; we will comment on this point again at the end
of Section~\ref{sec:resAnalytic}.
The only resonances that can contribute
at tree level are vector mesons, and isospin violation is provided
by the well-known mixing of the isovector $\rho^0$ and the isoscalar
$\omega$ state.  
We lay out the formalism for incorporating vector mesons,
their mixing and coupling to vector currents and nucleons 
in the following section before presenting analytical and numerical results 
for the isospin violating magnetic moments.
Contributions to the higher moments (electric and magnetic radii) 
will come as a benefit that allows us to estimate higher-order corrections 
to the leading chiral loop predictions.

\subsection{Formalism, Lagrangians\label{sec:resLagr}}

We here discuss the necessary ingredients to describe the interaction
of vector mesons with external currents as well as nucleons.  
We make use of the antisymmetric tensor formulation
that was shown to be particularly useful in the context of ChPT
and resonance saturation investigations~\cite{EGPdR,EGLPdR}.
Most of the literature is concerned with vector mesons in the context
of SU(3) ChPT; we will here only be concerned with the SU(2) subsystem
and adapt the formalism accordingly.  In SU(2), the antisymmetric tensor $V_{\mu\nu}$
is given in terms of $\rho$ and $\omega$ fields according to
\[ V_{\mu\nu} = \frac{1}{\sqrt{2}} \bigl( \vec\rho_{\mu\nu}\cdot\vec\tau
+ \omega_{\mu\nu} \bigr) ~.
\]
The free Lagrangian for $V_{\mu\nu}$ then takes the form
\beq
\Lagr_V = -\frac{1}{2} \langle D^\mu V_{\mu\nu} D_\rho V^{\rho\nu} \rangle
+ \frac{M_V^2}{4} \langle V_{\mu\nu} V^{\mu\nu} \rangle ~.
\label{Vkin}
\eeq
In particular, Eq.~\eqref{Vkin} leads to a common vector meson mass
$M_\rho=M_\omega=M_V$.  For numerical results, we will
use the mass of the $\rho$, $M_V=770$~MeV.
The coupling of the vector mesons to external vector (and axial vector) sources
is given by
\beq
\Lagr_{V\gamma} = \frac{F_V}{2\sqrt{2}} \langle V^{\mu\nu} {\cal F}^+_{\mu\nu} \rangle ~,
\label{Vgamma}
\eeq
where ${\cal F}^+_{\mu\nu}$ contains the electromagnetic field strength tensor
proportional to the \emph{quark} charge matrix $\mathcal{Q}=e/3\,\text{diag}(2,-1)$
(as opposed to the \emph{nucleon} charge matrix $Q$ used elsewhere in this text).
Eq.~\eqref{Vgamma} correctly reproduces the SU(3) relation for the 
vector meson decay constants $F_\rho = 3F_\omega = F_V$.
Phenomenologically, this relation is rather well
fulfilled, $F_\rho=152.5$~MeV, $F_\omega=45.7$~MeV.

Mixing of $\rho^0$ and $\omega$ has been discussed in a formalism
similar to the one presented here in Ref.~\cite{Urech}, albeit in the framework of SU(3).
Here, we perform an analogous construction for SU(2).
We are not interested in (isospin-conserving) quark mass renormalization
effects of the vector meson masses, but only in terms contributing
to $\rho$--$\omega$ mixing.  
One single such term can be constructed, 
\beq
\Lagr_{mix} = \frac{v_{mix}}{8}\langle V_{\mu\nu}\rangle
\langle \tilde V^{\mu\nu} \chi_+ \rangle ~,
\label{eq:Lmix}
\eeq
where we have once more made use of the notation 
$\tilde V_{\mu\nu} = V_{\mu\nu} - \frac{1}{2}\langle V_{\mu\nu}\rangle$.
If we match Eq.~\eqref{eq:Lmix} to the SU(3) result in
Ref.~\cite{Urech},
also invoking quark counting rules~\cite{QuarkMasses}, 
the mixing parameter can be identified with $v_{mix}=2M_V/B$.
In addition, there is a $\rho$--$\omega$ transition through an intermediate
photon via the interaction term Eq.~\eqref{Vgamma}.
Combining both effects, one finds the on-shell mixing amplitude
\beq
\Theta_{\rho\omega} = 
2M_V(m_u-m_d) + \frac{e^2 F_V^2}{3} ~.
\eeq
The latest analysis of experimental data~\cite{Ayse} yields
\beq 
\Theta_{\rho\omega} = (-3.75\pm 0.36)\times 10^{-3}~\text{GeV}^2 ~, 
\label{ThetaNum}
\eeq
well in agreement with earlier numbers~\cite{Urech}.

The coupling of vector mesons to baryons, formulated in the antisymmetric
tensor formalism, was discussed in Ref.~\cite{Borasoy}.  The relation
of the generic couplings used in that reference to the more conventional
vector and tensor couplings was given in Ref.~\cite{Kubis}.  
For a more compact presentation, we rewrite the terms from Ref.~\cite{Borasoy} 
in a SU(2) form, employing immediately standard couplings, which results in
\beq \begin{split}
\Lagr_{VN} &=
-\frac{M_V}{\sqrt{2}} \bar\Psi \biggl[ \frac{\sigma^{\mu\nu}}{4m_N}
\Bigl\{ g_\rho \kappa_\rho \tilde V_{\mu\nu} +
\frac{g_\omega \kappa_\omega}{2} \langle V_{\mu\nu} \rangle \Bigr\}  \\
&+ \frac{\gamma^\mu}{M_V^2} \Bigl\{
g_\rho \bigl[D^\nu,\tilde V_{\mu\nu}\bigr]
+\frac{g_\omega}{2} \bigl[D^\nu,\langle V_{\mu\nu}\rangle \bigr] \Bigr\} 
\biggr]\Psi ~.
\end{split} \eeq

We remark here that we totally neglect $\rho$--$\phi$ mixing in this analysis
(which would be present at least in an SU(3) extension).
Although the $\phi$--nucleon couplings are not as small as the Zweig rule
might lead one to expect, see e.g.\ Refs.~\cite{MMD,HM}, the mixing
is roughly an order of magnitude smaller than that between $\rho$ and
$\omega$~\cite{Bijnens,threemixing}, and therefore beyond the 
accuracy we aim to achieve by this model estimate.

\subsection{Analytical results\label{sec:resAnalytic}}

\begin{figure}
\includegraphics[height=2cm]{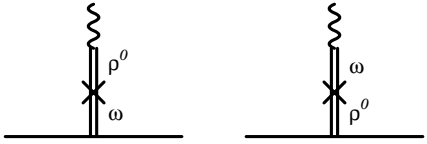}
\caption{\label{fig:mixing}
The two diagrams involving $\rho$--$\omega$ mixing
that contribute to the isospin breaking form factors.
The double lines denote vector meson propagators,
the cross represents the mixing vertex.
}
\end{figure} 
The $\rho$--$\omega$ mixing contributions to the isospin violating
Dirac and Pauli form factors can be derived from the two diagrams
depicted in Fig.~\ref{fig:mixing}.
They are given in terms of the various coupling
constants defined in the previous section as follows:
\beq \begin{split}
\Fv{1}(t) \,\Bigr|_{mix} &= -
\frac{g_\omega F_\rho\,\Theta_{\rho\omega}\,t}{M_V(M_V^2-t)^2} ~,\\
\Fs{1}(t) \,\Bigr|_{mix} &= -
\frac{g_\rho F_\omega\,\Theta_{\rho\omega}\,t}{M_V(M_V^2-t)^2} ~,\\
\Fv{2}(t) \,\Bigr|_{mix} &= -
\frac{g_\omega \kappa_\omega F_\rho\,M_V \,\Theta_{\rho\omega}}{(M_V^2-t)^2} ~,\\
\Fs{2}(t) \,\Bigr|_{mix} &= -
\frac{g_\rho \kappa_\rho F_\omega\,M_V \,\Theta_{\rho\omega}}{(M_V^2-t)^2} ~.
\end{split} \label{mixamp} \eeq
From these, one can easily derive the leading moments:
\beq \begin{split}
\kappa^{u,d}_{mix} &=
\bigl( g_\omega\kappa_\omega F_\rho - g_\rho\kappa_\rho F_\omega \bigr)
\frac{\Theta_{\rho\omega}}{M_V^3} ~, \\
\Rud{1}_{mix} &=
\bigl( g_\omega F_\rho- g_\rho F_\omega \bigr) 
\frac{\Theta_{\rho\omega}}{M_V^5} ~, \\
\Rud{2}_{mix} &=
\bigl( g_\omega\kappa_\omega F_\rho - g_\rho\kappa_\rho F_\omega \bigr)
\frac{2\Theta_{\rho\omega}}{M_V^5} ~.
\end{split} \label{leadud} \eeq
For the phenomenological discussion, we will be even more interested
in the leading moments of the Sachs form factors, which are given
in terms of the above as
\beq \begin{split}
\rud{E} &= \rud{1} + \frac{\kappa^{u,d}}{4m_N^2} ~,\\
\rud{M} &= \rud{1} + \rud{2} ~.
\end{split} \eeq

In light of the above results, we want to comment on the claim
made earlier that the saturation of coupling constants by
vector meson contributions might work better here than for the
isospin conserving form factors considered in Ref.~\cite{Kubis}.
The reason is that the additional propagator in the mixing case
leads to a higher power of vector meson masses in the denominators
of the leading moments, Eq.~\eqref{leadud}.
A heavier pair of isovector and isoscalar vector resonances 
sufficiently close to each other in mass to mix, 
e.g.\ the $\rho(1450)$ and the $\omega(1420)$~\cite{PDBook},
would yield contributions of the same \emph{form}
as Eq.~\eqref{leadud}, but
suppressed by a higher power of mass ratios $M_{V'}/M_V \approx 2$
than their unmixed contributions to the conventional form factors.  

Finally, we want to comment on the possibility of 
isospin violation other than through mixing.
In particular, it is possible to construct a mechanism for
``direct'' isospin breaking in the vector-meson--nucleon couplings
in analogy to Eq.~\eqref{L4}, 
\[
\Lagr_{VN}^{u,d} ~\propto~
\bar\Psi \,\sigma^{\mu\nu}
\Bigl\{ e_\rho \langle \tilde V_{\mu\nu} \tilde\chi_+ \rangle
+ e_\omega \langle V_{\mu\nu} \rangle \tilde\chi_+ \Bigr\}  
\Psi ~, 
\]
which results in the $\rho$ coupling as an isoscalar and 
the $\omega$ coupling as an isovector to the nucleons.  
(Analogous terms with the charge instead of the quark mass matrix
are easily written down.)
We disregard this possibility for the reason that the vector-meson--nucleon
coupling strengths are extracted from dispersive analyses on the assumption
of isospin symmetry (with the exception of $\rho$--$\omega$ mixing 
in the isovector spectral function), 
i.e.\ the $\omega$ couplings, for instance, are
identified as certain pole strengths in the isoscalar channel.
In this way, an isospin breaking $\rho$-nucleon coupling would just be
taken as part of the $\omega$ resonance, and vice versa.

\subsection{Numerical results\label{sec:resNumeric}}

The couplings of the vector mesons to nucleons are a rather delicate issue
and we prefer to rely as directly as possible on data rather than on models.
To this end,
we concentrate on values extracted from dispersive analyses of electromagnetic
form factors of the nucleon, 
and disregard values extracted from meson exchange
models of nucleon-nucleon scattering or pion-photo-/electroproduction
(see e.g.\ Refs.~\cite{Bonn,Nijm,Davidson,Pasquini}).
As it is well known that pure vector meson dominance does not yield
an adequate description of the isovector spectral function,
where the two-pion continuum leads to a significant enhancement 
on the left shoulder of the $\rho$ peak~\cite{BKM},
more recent analyses~\cite{MMD,MergellThesis,HM,BHM} make use of the full pion form factor
plus $\pi\pi\to N\bar N$ partial waves.  
In order to approximately disentangle the spectral function from Ref.~\cite{BHM}
into a non-resonant two-pion continuum plus a $\rho$ contribution,
we follow the method of Ref.~\cite{Kaiser} and add a Breit--Wigner parameterization
of the $\rho$ resonance to either the chiral one-loop or the two-loop~\cite{Kaiser}
representation of the two-pion cut contributions.  
This decomposition is model dependent, but probably adequate for a model
estimate of low-energy constants.
\begin{table*}
\caption{$\rho$--nucleon coupling constants and corresponding
form factor contributions.  ``1-loop'' and ``2-loop'' refer to the 
part of the non-resonant two-pion continuum that has been used to extract 
the $\rho$ couplings.  For details, see main text.\label{tab:rhoN}}
\renewcommand{\arraystretch}{1.6}
\begin{ruledtabular}
\begin{tabular}{cccc}
               &Ref.~\cite{MMD,MergellThesis}&Ref.~\cite{BHM} + 1-loop&Ref.~\cite{BHM} + 2-loop\\ 
\hline
$     g_\rho  $&  4.0       &  5.4      &  6.2         \\
$\kappa_\rho  $&  6.1       &  6.8      &  5.1         \\ 
\hline 
$10^2\times\ks$&  0.9       &  1.4      &  1.2         \\ 
$10^2\times\frac{\ks}{4m_N^2} ~\bigl[{\rm GeV}^{-2}\bigr]$
               &  0.3       &  0.4      &  0.3         \\ 
$10^2\times\rs{1} ~\bigl[{\rm GeV}^{-2}\bigr]$
               &  0.3       &  0.3      &  0.4         \\ 
$10^2\times\rs{2} ~\bigl[{\rm GeV}^{-2}\bigr]$
               &  3.1       &  4.6      &  4.0         \\ 
\end{tabular}
\end{ruledtabular}
\renewcommand{\arraystretch}{1.0}
\end{table*}  
The different values in Table~\ref{tab:rhoN} give a rather consistent picture
of the $\rho$-nucleon coupling constants on an accuracy level
of 20--30\%.  See Appendix~\ref{app:couplings} for the relations between
various coupling definitions.

\begin{table*}
\caption{$\omega$--nucleon coupling constants and corresponding
form factor contributions.  For the details, see main text.\label{tab:omegaN}}
\renewcommand{\arraystretch}{1.6}
\begin{ruledtabular}
\begin{tabular}{cccc}
               &Ref.~\cite{MMD}  &Ref.~\cite{HM}&Ref.~\cite{BHM2}\\ 
\hline
$     g_\omega$& $41.8$          & $43.0~~$     & $42.2~~$\\
$\kappa_\omega$& $\;-0.16$       & $~0.41$      & $~0.57$ \\ 
\hline
$10^2\times\kv$& $-0.8$          &  $2.2$       &  $3.0$  \\
$10^2\times\frac{\kv}{4m_N^2} ~\bigl[{\rm GeV}^{-2}\bigr]$
               & $-0.2$          &  $0.6$       &  $0.8$  \\
$10^2\times\rv{1} ~\bigl[{\rm GeV}^{-2}\bigr]$
               & $\phantom{-}8.8$&  $9.1$       &  $8.9$  \\
$10^2\times\rv{2} ~\bigl[{\rm GeV}^{-2}\bigr]$
               & $-2.8$          &  $7.5$       & $10.1~$ \\ 
\end{tabular}
\end{ruledtabular}
\renewcommand{\arraystretch}{1.0}
\end{table*}
The $\omega$ coupling constants are calculated from pure zero-width resonance
pole residues as found in dispersive analyses.  
The most noteworthy point about the numbers in Table~\ref{tab:omegaN} 
is the sign change in $\kappa_\omega$
in Refs.~\cite{HM,BHM2} as compared to Ref.~\cite{MMD} and other earlier analyses.
While the vector coupling $g_\omega$ seems to be determined consistently 
(although rather larger than what is inferred from $NN$-scattering~\cite{Bonn,Nijm}),
the tensor coupling $g_\omega \kappa_\omega$ is not at all, with not 
even the sign fixed.  
This uncertainty in $\kappa_\omega$ turns out to be by far the dominant
uncertainty in this analysis.

Considering the leading moments of $G_{E/M}^{u,d}$,
we observe the following:
\begin{enumerate}
\item $\kappa^{u,d}\,$:
The numbers in Tables~\ref{tab:rhoN}, \ref{tab:omegaN} suggest that,
as the small tensor coupling of the $\omega$ is relatively enhanced
by the larger $\rho$ decay constant,
$\ks$ and $\kv$
contribute numbers of similar size. 
The uncertainty in $\kappa_\omega$ completely dominates the
uncertainty of the estimate,
\beq
\kappa^{u,d}_{mix} = -0.020 \ldots +0.020 ~, \label{kappamix}
\eeq
where we have also included the error in the determination
of the mixing angle in Eq.~\eqref{ThetaNum}.
The negative value corresponds to the most recent dispersive 
analysis~\cite{BHM2}.
\item $\rud{E}\,$:
The large vector coupling of the $\omega$, enhanced by the $\rho$ decay constant,
completely dominates this quantity, such that one has
\[
\Rud{E}_{mix} \approx - \Rv{E}_{mix} ~~.
\]
Foldy terms hardly play a role, such that even the large uncertainty
in $\kappa_\omega$ does not spoil the prediction.  We find
\beq
\Rud{E}_{mix} = 
-\bigl[0.07\ldots 0.10\bigr]\,{\rm GeV}^{-2} ~,
\label{rEmix}
\eeq
where the range includes the uncertainty in $\Theta_{\rho\omega}$.

\item $\rud{M}\,$:
Again, the $\omega$ couplings to the nucleon yield the largest contributions.
The uncertainty is dominated by the uncertainty in $\kappa_\omega$, leading to a range
\beq
\Rud{M}_{mix} =
-\bigl[0.03 \ldots 0.15\bigr]\,{\rm GeV}^{-2} ~.
\label{rMmix}
\eeq
The recent analysis~\cite{BHM2} suggests the values larger in absolute magnitude
within that range.

\end{enumerate}

\subsection{Discussion\label{sec:Discussion}}

The above estimates for $\kappa^{u,d}$, $\rud{E}$,
$\rud{M}$ correspond to low-energy constants that 
enter the chiral representations of the corresponding form factors 
at $\Order(p^4)$, $\Order(p^5)$, $\Order(p^6)$, respectively.  
As we have only calculated $G_E^{u,d}$ up to $\Order(p^4)$
and $G_M^{u,d}$ up to $\Order(p^5)$, the estimate for $\kappa^{u,d}$
is the only one that is strictly needed for numerical results.
However, in order to test the potential size of higher-order corrections,
it may still be very useful to compare the estimates of the previous
subsection to the chiral loop contributions.

One downside of the resonance saturation method is that, 
if we want to identify some resonance contribution with
the finite part of a chiral coupling constant, it is not obvious
at what \emph{scale} this identification is to be made.  
It seems to be common wisdom, though, that this scale ought to be
roughly at the resonance mass; we will therefore choose $\lambda=M_\rho$.
We decompose Eq.~\eqref{eq:G0} according to
\beq
\kappa^{u,d} = \kappa^{u,d}_\chi(M_\rho) + \kappa^{u,d}_{CT}(M_\rho) 
\eeq
and set $\kappa^{u,d}_{CT}(M_\rho) = \kappa^{u,d}_{mix}$.  
Numerically, we find from Eq.~\eqref{eq:G0} for the chiral part
\beq
\kappa^{u,d}_\chi(M_\rho) = 0.025 ~.
\eeq
If we vary the saturation scale, $\lambda$, in the range $0.5\ldots 1.0$~GeV, we find
$\kappa^{u,d}_\chi=0.020\ldots 0.028$,
which may serve as an indicator of the uncertainty of the 
resonance saturation method as such.
These values are all of the same order of magnitude as $\kappa^{u,d}_{mix}$
in Eq.~\eqref{kappamix}.  
Even within the large error range for $\kappa^{u,d}_{mix}$, 
however, we predict $\kappa^{u,d}$ to be \emph{positive},
\beq
\kappa^{u,d} =   0.005 \ldots 0.045 ~.
\eeq
The most recent values for the coupling constants, with a negative
$\kappa^{u,d}_{mix}$, lead to a substantial cancellation between loop effects
and counterterm contributions, and altogether to a very small total $\kappa^{u,d}$.

\begin{figure}
\includegraphics[width=8.5cm]{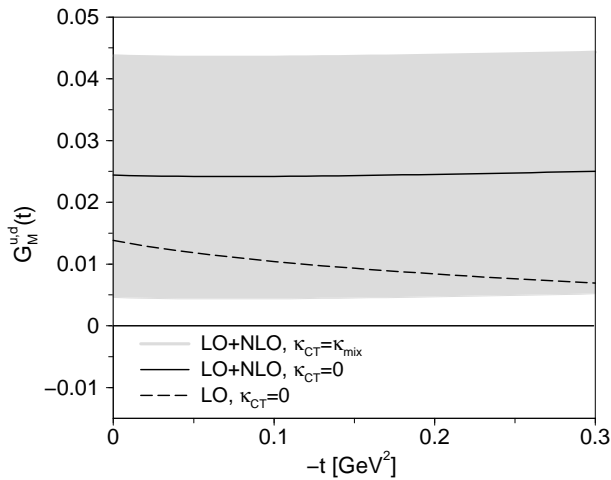}
\caption{\label{fig:GM1}
The form factor $G_M^{u,d}(t)$. 
The dashed line is the LO chiral prediction,
the full line includes the NLO order corrections,
both with $\kappa_{CT}^{u,d}(M_\rho)=0$.
The grey band is the chiral NLO representation with 
$\kappa_{CT}^{u,d}(M_\rho)=-0.020\ldots+0.020$.
}\end{figure}
For the magnetic radius term, the two leading chiral
contributions are unambiguously given in terms of loop effects.
Evaluating Eq.~\eqref{eq:rMchiral} numerically, 
we find the chiral prediction at NLO to result in
\beq
\Rud{M}_\chi= \bigl(0.05-0.04\bigr)\,{\rm GeV}^{-2} 
= 0.01\,{\rm GeV}^{-2} ~.
\eeq
Fig.~\ref{fig:GM1} shows the purely chiral NLO representation of $G_M^{u,d}(t)$
in the range $0\leq -t \leq 0.3\,$GeV$^2$,
together with the range of counterterm values as estimated
from Eq.~\eqref{kappamix}, shown as a grey band.  
For comparison, the LO and NLO representations are also shown for
$\kappa_{CT}^{u,d}(M_\rho)=0$.
Although the uncertainty is sizeable, $G_M^{u,d}(t)$ is predicted
to be positive and smaller than 0.05 in this range.  
Due to the substantial cancellation between chiral LO and NLO contributions
to the magnetic radius, the $t$-dependence is very weak.  Also,
the curvature induced by chiral loop effects is minimal.

\begin{figure}
\includegraphics[width=8.5cm]{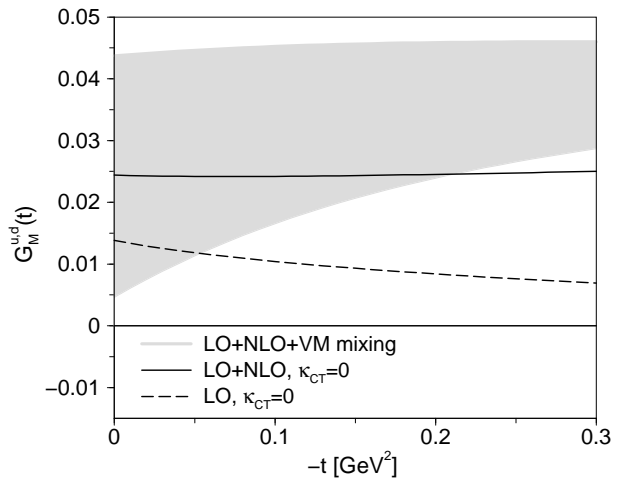}
\caption{\label{fig:GM2}
The form factor $G_M^{u,d}(t)$, notation as in Fig.~\ref{fig:GM1}.
The grey band now includes the full $t$-dependence of the
mixing amplitude.  For details, see main text.
}\end{figure}
Even if the large cancellation between the LO and NLO
terms for the magnetic radius is considered accidental, 
the (formally next-to-next-to-leading)
vector meson contribution in Eq.~\eqref{rMmix} is seen to be 
at least of the same order of magnitude, potentially larger than both.
The vector meson mixing contributions lead to a sign change
in the radius as compared to the chiral prediction.
Together with a positive magnetic moment, this means the form factors
increase in absolute magnitude for non-vanishing virtuality
in electron scattering experiments, $t<0$. 
This is shown in Fig.~\ref{fig:GM2}, where, in comparison to Fig.~\ref{fig:GM1},
we have replaced the low-energy constant contribution
$\kappa^{u,d}_{CT}(M_\rho) = \kappa^{u,d}_{mix}$
by the full $t$-dependence of the mixing amplitude, Eq.~\eqref{mixamp}.
As the complete mixing amplitude contains no more
parameters than the constant at $t=0$, the uncertainty band 
can even get narrower:  the upper boundary of the band (given 
by the vector meson parameters from Ref.~\cite{MMD}) changes very little 
and stays below $0.05$, while the lower boundary 
(given essentially by the parameters from Ref.~\cite{BHM2})
rises with $-t$.

We want to emphasize that this combined chiral plus vector meson mixing
representation as shown in Fig.~\ref{fig:GM2} goes beyond strict 
effective field theory.  We believe however that the additional
$t$-dependence of the mixing contributions provides a good estimate
of the most important higher-order terms that go beyond our chiral 
calculation, for the following reasons: 
the strongest $t$-dependence from pion loops has to correspond 
to cuts in the low-energy region, and among these, 
two-pion cuts are certainly the most prominent 
(see e.g.\ Ref.~\cite{BKM} on three-pion cut contributions to 
nucleon form factors); but we have already calculated
the two-pion cuts up to NLO, and we do not expect 
higher-order corrections to these to be unreasonably large
(witness Ref.~\cite{Kaiser} for two-loop corrections to the two-pion continuum).
We therefore find it reasonable to expect low-energy constants
(corresponding to resonance physics) to dominate the missing pieces
beyond the NLO chiral representation.
We have argued earlier that we expect resonance contributions
beyond $\rho$--$\omega$ mixing to be minor corrections, and consider their 
possible impact well-covered by the uncertainty bands in Figs.~\ref{fig:GM1}, \ref{fig:GM2}.
Note finally that such a combined approach was shown to work rather
well for the isospin conserving form factors in Refs.~\cite{Kubis,KubisHyperon}.

We now turn to the electric form factor $G_E^{u,d}(t)$.
It can only be predicted unambiguously to leading order in ChPT, 
where one has $G_E^{u,d}(t)=F_1^{u,d}(t)$.
\begin{figure}
\includegraphics[width=8.5cm]{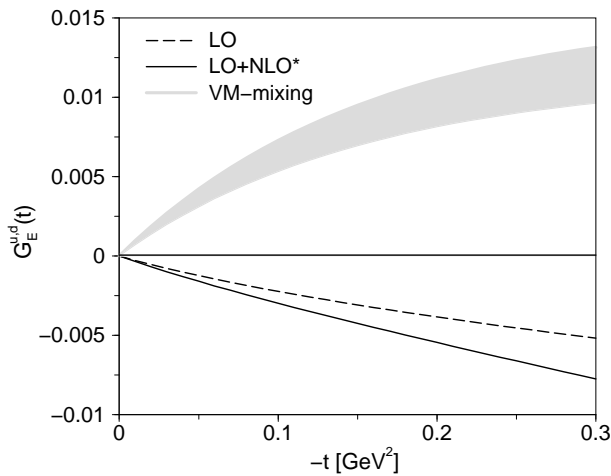}
\caption{\label{fig:GE1}
Contributions to the form factor $G_E^{u,d}(t)$. 
The full line is the LO chiral prediction,
the dashed line contains partial NLO order corrections
as discussed in the main text.  
The grey band is the vector meson mixing contribution.
}\end{figure}
This is shown as the dashed line in Fig.~\ref{fig:GE1}.
A \emph{partial} higher-order correction is given by 
the chiral contributions to $t/4m_N^2\times F_2^{u,d}(t)$, see Eq.~\eqref{Sachs},
which is added to the leading-order expression
for the full line in Fig.~\ref{fig:GE1} and can be seen there
to be also numerically subleading.

The leading chiral prediction for the electric radius term is,
from Eq.~\eqref{eq:F1t} and Appendix~\ref{app:loopfunctions},
\beq
\Rud{E}_\chi = 
\frac{5\pi\,C}{6M_\pi m_N} +\Order\bigl(M_\pi^0\Delta m \bigr)~. 
\eeq
Numerically, this amounts to 
\beq
\Rud{E}_\chi= 0.03\,{\rm GeV}^{-2} ~,
\eeq
which demonstrates that the vector meson contribution, see Eq.~\eqref{rEmix},
is numerically dominant, albeit formally subleading.
It again leads to a sign change compared to the leading chiral prediction.
This can also be seen from Fig.~\ref{fig:GE1}, where the full $\rho$--$\omega$
mixing amplitude is depicted as a grey band, corresponding to the 
range of vector meson coupling constants yielding the electric radius range in
Eq.~\eqref{rEmix}.  
Compared to $G_M^{u,d}(t)$,
the band is better constrained as the particularly controversial
coupling $\kappa^\omega$ plays no major role in the electric form factor.
\begin{figure}
\includegraphics[width=8.5cm]{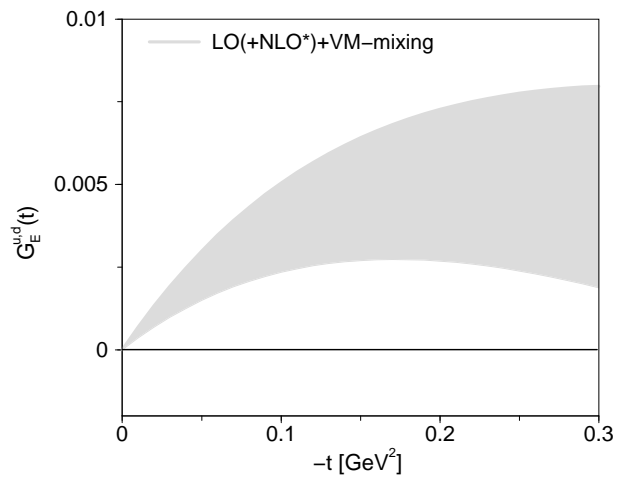}
\caption{\label{fig:GE2}
The complete form factor $G_E^{u,d}(t)$.
The band combines uncertainties in the vector meson mixing amplitude
as well as from higher-order chiral corrections.
For details see main text.
}\end{figure}
Chiral and vector meson contributions are combined in Fig.~\ref{fig:GE2}, 
where we have taken the partial NLO chiral contributions described above
as an uncertainty for higher-order loop corrections. Errors
from both sources were added linearly.  We consider this band a 
conservative estimate.
Due to the bigger mixing amplitude, the 
total form factor is positive, but remains small ($<0.01$) 
and therefore well-constrained in the whole momentum transfer range considered.

\begin{table*}
\caption{Comparison of selected experimental measurements of strange form factors
to the results of this work.  The column ``electric/magnetic'' shows the linear
combination of strange electric and magnetic form factors as well as the 
momentum transfer in the respective kinematics, the column ``$G^s$'' the corresponding
result.  The column ``$G^{u,d}$'' gives the range of the isospin breaking
form factor contributions for the same linear combination, in the same kinematics,
as predicted in this work.
\label{tab:strange}}
\renewcommand{\arraystretch}{1.6}
\begin{ruledtabular}
\begin{tabular}{cccc}
experiment     & electric/magnetic & $G^s$    & $G^{u,d}$   
\\ \hline
SAMPLE~\cite{SAMPLE}  & $G_M\bigl(-0.1\,{\rm GeV}^2\bigr)$ & 
$0.37 \pm 0.20 \pm 0.26 \pm 0.07$     & $0.02 \ldots 0.05$
\\ 
A4~\cite{A4b}         & $\bigl[G_E+0.106\,G_M\bigr]\bigl(-0.108\,{\rm GeV}^2\bigr)$ & 
$0.071 \pm 0.036$                     & $0.004 \ldots 0.010$
\\ 
HAPPEX~\cite{HAPPEXb} & $\bigl[G_E+0.080\,G_M\bigr]\bigl(-0.099\,{\rm GeV}^2\bigr)$ & 
$0.030 \pm 0.025 \pm 0.006 \pm 0.012$ & $0.004 \ldots 0.009$
\\
\end{tabular}
\end{ruledtabular}
\renewcommand{\arraystretch}{1.0}
\end{table*}
In order to put these numbers into perspective concerning the strangeness form 
factor measurements, we want to compare them to some experimental results on the latter.
Ref.~\cite{Young} is a recent attempt to combine
all available world data on parity violating electron scattering
and perform a best fit for the leading strangeness moments.
The fit including only leading-order moments (e.g.\ no strange magnetic radius)
yields
\beq \begin{split}
\kappa^s &= 0.12 \pm 0.55 \pm 0.07 ~, \\
\rho_E^s &= \bigl(-0.06 \pm 0.41 \mp 0.00\bigr)\,{\rm GeV}^{-2} ~, 
\end{split} \label{strangestatus}\eeq
while a fit allowing for next-to-leading-order moments results in
$\rho_M^s = (0.7 \pm 6.8)\,$GeV$^{-2}$,
i.e.\ the data are not sufficiently accurate yet to pin down the magnetic
radius to reasonable accuracy.
So while the central values of Eq.~\eqref{strangestatus} are already of comparable
magnitude to $\kappa^{u,d}$, $\rud{E}$,
the isospin violating moments are still, by a factor of 6--10, smaller than
the combined uncertainties on the strangeness moments.

In Table~\ref{tab:strange}, we also compare to a few selected individual 
experimental numbers on strangeness form factors~\cite{SAMPLE,HAPPEXb,A4b}.
We contrast those results with bands for the isospin violating form factors
in the same kinematics, i.e.\ for the same combination of electric and
magnetic, and the same momentum transfer $t$. 
The error band corresponds to a worst-case combination of
the bands discussed in connection with Figs.~\ref{fig:GM2}, \ref{fig:GE2}.
The conclusion is similar here:  the uncertainty from isospin breaking 
is still smaller than the overall experimental error, but is relevant
as measurements become more and more precise.

\section{Summary and conclusions\label{sec:Conclusions}}

In this paper, we have re-investigated the isospin breaking vector
form factors of the nucleon that are increasingly becoming a necessary ingredient
to the precise extractions of the strange vector form factors.
Our findings can be summarized as follows:
\begin{enumerate}
\item
We  calculate the isospin breaking electric and magnetic form factors
$G_{E/M}^{u,d}(t)$ to leading and next-to-leading order in 
two-flavor chiral perturbation theory, correcting some errors in Ref.~\cite{LewisMobed}.
Up to this order of accuracy, all loop effects are due to the
proton-neutron mass difference.
Only one combination of unknown low-energy constants enters
the chiral representation of these form factors, namely a contribution
to the isospin breaking magnetic moment.  
Leading and next-to-leading-order contributions to the magnetic radius
are seen to cancel to a large extent, leading to a very weak momentum 
dependence.
\item
In order to estimate the missing counterterms, we employ the method
of resonance saturation.
We provide analytic expressions for the isospin breaking magnetic moment
as well as for (formally subleading) electric and magnetic radii
in terms of various vector meson coupling constants and the
$\rho$--$\omega$ mixing angle.  
While most of these phenomenological parameters are known to apt precision,
by far the most uncertain coupling is the $\omega$--nucleon tensor coupling, 
which dominates the rather broad error band 
in our prediction for the isospin breaking magnetic form factor.
\item
For the electric and magnetic radii where counter\-terms only come in
at subleading orders, we find that
the combination of the relatively small nucleon mass
difference, which is responsible for the chiral loop contributions,
and the strong vector-meson--nucleon couplings, which numerically enhance 
the effect of low-energy constants, tends to upset the hierarchy suggested
by chiral power counting.  
The vector meson mixing contributions lead to sign changes in the radii
compared to the purely chiral predictions.
\item 
We combine chiral loop contributions and the full vector meson mixing amplitudes
in a phenomenological approach, adding the various uncertainties
to produce conservative error bands.  We find that $G_{E/M}^{u,d}(t)$ 
are both positive in the momentum transfer range $0\leq -t \leq 0.3\,$GeV$^2$,
with approximate upper limits $G_M^{u,d}(t)<0.05$,  $G_E^{u,d}(t)<0.01$.
\end{enumerate}
In order to sharpen our findings and make the prediction for 
the isospin breaking vector form factors even more stringent, 
the predominant task would be to improve on our knowledge of the
$\omega$--nucleon tensor coupling, if feasible.  
Higher-order chiral calculations would be extremely tedious
as they would involve two-loop diagrams with pions and photons, 
and are unlikely to be of high numerical relevance compared to 
the resonance physics considered in this article.

\begin{acknowledgments}
We are grateful to Nader Mobed for helpful communications related to the
heavy-baryon ChPT calculation, and
we would like to thank Maxim Belushkin, Hans-Werner Hammer, and Ulf-G.\ Mei{\ss}ner
for useful discussions on dispersive analyses of the nucleon form factors.
Partial financial support under the EU Integrated Infrastructure
Initiative Hadron Physics Project (contract number RII3-CT-2004-506078)
and DFG (SFB/TR 16, ``Subnuclear Structure of Matter'') is gratefully
acknowledged.
Furthermore, this work was supported by 
the Natural Sciences and Engineering Research Council of Canada
and the Canada Research Chairs Program.
One of us (R.L.) would like to thank the theory group of the 
IKP at the Forschungszentrum J\"ulich
for support and hospitality during parts of this work.

\end{acknowledgments}

\appendix

\section{Chiral background}

\subsection{Loop functions\label{app:loopfunctions}}

Here we spell out the explicit forms of the loop functions used
in Section~\ref{sec:chiralff}:
\beq \begin{split}
\bar\gamma_0(t)
 &= \frac{1}{16\pi}\biggl(\frac{2M_\pi}{\sqrt{-t}}
     \arctan \frac{\sqrt{-t}}{2M_\pi}-1\biggr) ~,\\
\bar\gamma_3(t)
 &= \frac{1}{16\pi}\biggl(\frac{4M_\pi^2-t}{4M_\pi\sqrt{-t}} 
                \arctan\frac{\sqrt{-t}}{2M_\pi} -\frac{1}{2}\biggr) ~,~~\\
\xi(t)
 &= -\frac{1}{16\pi^2}\biggl(\sigma \log\frac{\sigma+1}{\sigma-1} - 2 \biggr) ~, 
\end{split} \label{eq:loopfunctions}\eeq
where $\sigma=\sqrt{1-{4M_\pi^2}/{t}}$.  
$\bar\gamma_{0/3}(t)$ are linked to the loop functions $\gamma_{0/3}(t)$ 
used in Ref.~\cite{BKMreview} via
\[
\bar\gamma_0(t) = M_\pi\bigl(\gamma_0(t)-\gamma_0(0)\bigr) ~,~~
\bar\gamma_3(t) = M_\pi^{-1}\bigl(\gamma_3(t)-\gamma_3(0)\bigr) ~.
\]
The leading $t$-dependence of these functions is given by
\beq \begin{split}
\bar\gamma_0(t) &= \frac{t}{192\pi M_\pi^2} + \Order(t^2) ~, \\
\bar\gamma_3(t) &= - \frac{t}{192\pi M_\pi^2} + \Order(t^2) ,\\
\xi(t) &= \frac{t}{96\pi^2M_\pi^2} + \Order(t^2) ~.
\end{split} \eeq
We note that the functions in Eq.~\eqref{eq:loopfunctions} 
are the heavy-baryon loop functions
that were reproduced from the infrared regularized loops by
strict expansion in chiral powers.  
As we are only interested in the form factors in the space-like 
region, we disregard the complications ensuing from the anomalous
threshold of the triangle diagram.  We want to emphasize, though, 
that these representations do not reproduce the correct threshold
behavior of the spectral functions.

\subsection{Separate diagrams}

For completeness, 
we show here the various contributions to $\Fv{1/2}(0)$, $\Fs{1/2}(0)$
from the diagrams displayed in Fig.~\ref{fig:pionloops}.  
Of course $\Fv{1}(0)=\Fs{1}(0)=0$ in the sum of all diagrams is
a necessary requirement.  
The prefactor $C$ is as defined in Eq.~\eqref{Cdef}.
\begin{align}
\Fv{1}(0)\bigl[ \text{(a)} \bigr]
&= -2\Fv{1}\bigl[ \text{(b)} \bigr]
= -2\Fv{1}\bigl[ \text{(c)} \bigr]
= \frac{6\pi M_\pi}{m_N} \,C 
~,\no\\
\Fs{1}\bigl[ \text{(b)} \bigr]
&=-\Fs{1}\bigl[ \text{(c)} \bigr]
=\frac{3\pi M_\pi}{m_N} \,C ~, 
\end{align}
\begin{align}
\Fv{2}(0)\bigl[ \text{(a)} \bigr]
&= 4\biggl[ 1+2\log\frac{M_\pi}{\lambda} 
-\frac{4\pi M_\pi}{m_N} \biggr] \,C 
~, \no \\
\Fv{2}\bigl[ \text{(b)} \bigr]
&=-\Fs{2}\bigl[ \text{(b)} \bigr]
= \frac{4\pi M_\pi}{m_N} \,C 
~, \no \\ 
\Fv{2}\bigl[ \text{(d)} \bigr]
&= -\frac{1}{3}\Fv{2}\bigl[ \text{(e)} \bigr]
= \frac{\pi (\kappa_p-\kappa_n)M_\pi}{m_N} \,C 
~, \no\\
\Fs{2}\bigl[ \text{(d)} \bigr]
&= \frac{1}{3}\Fs{2}\bigl[ \text{(e)} \bigr]
= -\frac{\pi(\kappa_p+\kappa_n) M_\pi}{m_N} \,C 
~. 
\end{align}

\section{Vector meson coupling constants\label{app:couplings}}

In this appendix, we briefly spell out how to relate the vector-meson--nucleon
coupling constants used in this paper to those in Refs.~\cite{Kaiser,MMD,HM,BHM2}.

In Ref.~\cite{Kaiser}, the strength of the $\rho$ contributions to the
isovector spectral functions of the Sachs form factors is parameterized 
in terms of two coupling constants $b_{E,M}$ that are given in terms 
of $g_\rho$, $\kappa_\rho$ according to
\beq
b_E = \frac{g_\rho F_\rho}{2M_\rho} \biggl(1+\frac{\kappa_\rho M_\rho^2}{4m_N^2}\biggr) ~,
\quad
b_M = \frac{g_\rho F_\rho}{2M_\rho} \bigl(1+\kappa_\rho\bigr) ~.
\label{bEbM}
\eeq
In Ref.~\cite{Kaiser}, the numerical values $b_E=1.0$, $b_M=3.6$ were extracted.
Using the updated empirical isovector spectral function from Ref.~\cite{BHM}, 
we find the slightly shifted numbers $b_E=1.1$, $b_M=3.7$, while fitting
the phenomenological $\rho$ contribution with just the one-loop two-pion 
continuum leads to $b_E=1.2$, $b_M=4.2$.  These numbers, together with
Eq.~\eqref{bEbM}, feed into Table~\ref{tab:rhoN}.

The $\omega$ coupling constants are given as pole residues $a_{1,2}^\omega$ 
in the isoscalar spectral functions of Dirac and Pauli form factors
in Refs.~\cite{MMD,HM,BHM2}.  The coupling constants $g_\omega$, $\kappa_\omega$ 
can be calculated from these as
\beq
g_\omega = \frac{2a_1^\omega}{F_\omega M_\omega} ~, \quad
\kappa_\omega = \frac{a_2^\omega}{a_1^\omega} ~.
\label{aiomega}
\eeq
The pole residues from Ref.~\cite{BHM2} used for Table~\ref{tab:omegaN} are
$a_1^\omega = 0.752$ GeV$^2$, $a_2^\omega=0.425$ GeV$^2$.

\bibliography{ffisobreak}

\end{document}